\documentclass[conference]{IEEEtran}
\usepackage{graphicx}
\usepackage{amsmath,bbm,epsfig,amssymb,amsfonts, amstext, verbatim,amsopn,cite,subfigure,multirow,multicol,lipsum}
\usepackage{balance}
\usepackage{url}
\usepackage{amsfonts}
\usepackage{epsfig}
\usepackage{setspace}
\usepackage{stmaryrd}
\usepackage{psfrag}	
\usepackage{multirow}
\usepackage{float}
\usepackage[process=auto]{pstool}
\usepackage{etoolbox}
\usepackage{algorithm}
\usepackage{algorithmic}
\usepackage{hyperref}

\usepackage{pifont}
\allowdisplaybreaks

\newtheorem{lem}{Lemma}
\newtheorem{remk}{Remark}

\newtheorem{prop}{Proposition}
\newtheorem{corol}{Corollary}

\newtoggle{ConfVersion}

\togglefalse{ConfVersion}

\iftoggle{ConfVersion}{%
}{%
}

\EndPreamble
\begin{document}

\title{Non-Orthogonal Multiple Access for \\ FSO Backhauling
\vspace{-0.3cm}}
\author{Marzieh Najafi$^{\dag}$, Vahid Jamali$^{\dag}$,  Panagiotis D. Diamantoulakis$^{\ddag}$, George K. Karagiannidis$^{\ddag}$, and Robert Schober$^{\dag}$\\  
$^{\dag}$Friedrich-Alexander University of Erlangen-Nuremberg, Germany \\
$^{\ddag}$Aristotle University of Thessaloniki, Greece
\vspace{-0.2cm}}

\maketitle
\begin{abstract}
We consider a free space optical (FSO) backhauling system which consists of two base stations (BSs) and one central unit (CU). We propose to employ non-orthogonal multiple access (NOMA) for FSO backhauling where both BSs transmit at the same time and in the same frequency band to the same photodetector at the CU. We develop a dynamic NOMA scheme which determines the optimal decoding order as a function of the channel state information at the CU and the quality of service  requirements  of the BSs, such that the outage probabilities of both BSs are jointly minimized. Moreover, we analyze the performance of the proposed NOMA scheme in terms of the outage probability over Gamma-Gamma FSO turbulence channels. We further derive closed-form expressions for the  outage probability for the high signal-to-noise ratio regime. Our simulation results confirm the analytical derivations and reveal that the proposed dynamic NOMA scheme significantly outperforms orthogonal transmission and existing NOMA schemes. 
\end{abstract}

\section{Introduction}
The ever-growing demand for higher data rates in the last few decades places tremendous pressure on existing networks and has become the main 
challenge and research focus for the design of the next generation  wireless communication systems \cite{Survey_Cheng}. 
 Therefore, larger bandwidths and more spectrally efficient transmission schemes are needed to support the data rate requirements of fifth generation (5G) wireless systems. Free space optical (FSO) systems are considered to be promising candidates for backhauling in 5G systems since they offer huge bandwidths \cite{Survey_Cheng,Optic_Tbit}. In addition, FSO systems are inherently secure and energy efficient \cite{Survey_Cheng}. 
 

Non-orthogonal multiple access (NOMA) has been proposed for 5G systems to increase the spectral efficiency of the radio access network compared to conventional orthogonal multiple access (OMA) \cite{NOMA_Survey_Robert,NOMA_George,SIC_NOMA}, by multiplexing more than one user on the same time-frequency resource  \cite{NOMA_Survey_Robert,NOMA_George}. Moreover, NOMA provides massive connectivity and low latency which are other important requirements for 5G systems \cite{NOMA_Survey_Robert,SIC_NOMA}. It is also noted that the special case of two-user downlink NOMA, termed as multi-user superposition transmission (MUST), has been included in the 3rd Generation Partnership Project (3GPP) Long Term Evolution Advanced (LTE-A) standard \cite{3GPP_NOMA}. 
Since in NOMA the superposition of several users' messages is received, successive interference cancellation (SIC) is adopted at the receiver to separate and decode the users' messages \cite{NOMA_Survey_Robert,SIC_NOMA}. Note that for  downlink NOMA, the decoding order at the receivers is fixed whereas in  uplink NOMA, different decoding orders are possible \cite{NOMA_Survey_Robert}. In particular, in the literature, the decoding order of the users is assumed to be either a priori fixed \cite{FixedSIC} or dynamic where,  for example, the user with higher instantaneous received signal power is decoded first  \cite{Dynamic_SIC_NOMA}. Moreover, for performance analysis of NOMA systems, three assumptions are commonly made regarding SIC reliability, namely the \textit{perfect, imperfect}, and \textit{worst-case SIC} assumptions. Under the {perfect SIC} assumption, it is assumed that while decoding the message of a specific user, the interference from other users with higher decoding order has been perfectly cancelled \cite{SIC_NOMA}. Under the {imperfect SIC} assumption, if decoding the message of a user is unsuccessful, its effect on the signal of other users with lower decoding orders is taken into account as interference \cite{SIC_NOMA,NOMA_Survey_Robert}.  Under the {worst-case SIC} assumption, if decoding the message of a user is unsuccessful, the decoding of the messages of all users with lower decoding orders is assumed to be unsuccessful  \cite{SIC_NOMA,Dynamic_SIC_NOMA}. Note that imperfect SIC is a more realistic assumption than perfect and worst-case SIC; nevertheless, the latter two assumptions are widely adopted in the literature due to their mathematical tractability.

In this paper, we investigate NOMA for FSO backhauling in 5G systems. In particular, we assume that two base stations (BSs) simultaneously send their data to a central unit (CU) over FSO backhaul links. The main motivation for combining FSO and NOMA is the objective to exploit the benefits of  both the high data rates enabled by FSO  and the high spectral efficiency introduced by NOMA. Another motivation is that NOMA usually performs well when the powers received from different transmitters are quite different \cite{NOMA_Survey_Robert}. This is typically the case for FSO systems since the path loss of FSO links is higher than that of radio frequency (RF) links, particularly in adverse weather conditions, such as haze and fog. In this paper, we propose a dynamic NOMA scheme which determines the optimal decoding order, such that the outage probabilities of both BSs are  jointly minimized. We prove that the proposed decoding order strategy is optimal under the perfect, imperfect, and worst-case SIC assumptions. Moreover, we analyze the system performance in terms of the outage probability of the BSs over FSO Gamma-Gamma (G-G) turbulence channels. We further derive closed-form expressions for the outage probability in the high signal-to-noise ratio (SNR) regime. Our simulation results confirm the analytical derivations and reveal that the proposed dynamic NOMA scheme significantly outperforms the fixed and dynamic NOMA schemes from the literature.

We note that so far, NOMA has been mainly considered for radio access, i.e., uplink or downlink transmission in a system comprising  several users and a BS, in RF  \cite{NOMA_Survey_Robert,NOMA_George,SIC_NOMA}, millimeter wave \cite{mmWNOMA}, and visible light communication systems \cite{VLC_NOMA_George}. On the contrary, in this paper, we consider NOMA for wireless backhauling of two BSs to a CU. Moreover, except for the recent paper \cite{arXiv_NOMA_FSO}, NOMA has not been considered previously for use in FSO systems. Furthermore, unlike this paper,  the authors of \cite{arXiv_NOMA_FSO} employ the well-known dynamic SIC scheme which sorts the users for decoding according to their channel qualities \cite{Dynamic_SIC_NOMA}. In other words, the dynamic NOMA in \cite{arXiv_NOMA_FSO} exploits only the channel state information (CSI)  to determine the  decoding order whereas the proposed optimal NOMA scheme takes both the CSI and the quality of service (QoS) requirements of the BSs into account when determining the \textit{optimal} decoding order. Therefore, the performance analysis in this paper is different from the one carried out in \cite{arXiv_NOMA_FSO}. Furthermore, we provide a high SNR performance analysis which was not offered in \cite{arXiv_NOMA_FSO}. We derive conditions for the BSs' QoS requirements which determine whether or not the outage probability exhibits an outage floor. Finally, we show that the proposed optimal dynamic scheme outperforms the dynamic scheme in \cite{arXiv_NOMA_FSO}.

\section{System and Channel Models}\label{SysMod}

In this section, we present the system and channel models adopted in this paper. 

\subsection{System Model}
We consider an FSO backhauling system where two BSs denoted by BS$_1$ and BS$_2$ wish to communicate with a single CU over FSO links. We assume that the BSs and the CU are fixed nodes mounted on top of buildings where a line-of-sight  between the BSs and CU is available. The CU is equipped with a single photodetector (PD) with a circular detection aperture of radius $r$.  Moreover, each BS has a single aperture directed towards the PD at the CU.  We study NOMA for FSO backhauling where the two BSs transmit their signals to the CU at the same time and in the same frequency band. The CU employs SIC to decode the BSs' signals.  We propose dynamic-order decoding where the decoding order at the CU is a function of the FSO channel conditions and the  QoS requirements of the BSs. In particular, we choose the decoding order, such that the outage probability of the BSs is minimized. 
%
%
%
\subsection{Channel Model}
 We assume an intensity modulated direct detection (IM/DD) FSO system. Particularly, after removing the ambient background light intensity, the signal intensity detected at the CU is modelled as
\begin{IEEEeqnarray}{lll}\label{Eq:received_signal}
y=h_{1}x_1+h_{2}x_2+n,
\end{IEEEeqnarray}
where $x_1$ and $x_2$ are the optical signals transmitted by BS$_1$ and BS$_2$, respectively, and $n$ is zero-mean real-valued additive white Gaussian shot noise (AWGN) with variance $\delta^2_n$ caused by ambient light. Moreover, $h_1$ and $h_2$ are the independent positive real-valued channel coefficients from BS$_1$ and BS$_2$ to the CU, respectively. Furthermore, we assume an average optical power constraint $\mathbb{E}\{x_i\}\leq P_i$, where $\mathbb{E}\{\cdot\}$ denotes expectation. 

In the following, we present the model for $h_i,i=1,2$, adopted in this paper. In particular, $h_i$ is affected by several factors and can be modelled as follows 
\begin{IEEEeqnarray}{lll}\label{Eq:channel}
h_i=\rho\bar{h}_i\hat{h}_i\tilde{h}_{i},
\end{IEEEeqnarray}
where $\rho$ is the responsivity of the PD and $\bar{h}_i$, $\hat{h}_i$, and $\tilde{h}_i$ are the path loss, geometric loss, and  atmospheric turbulence, respectively. The models adopted for $\bar{h}_i$, $\hat{h}_i$, and $\tilde{h}_{i}$ are presented in the following. 

\subsubsection{Path Loss}
$\bar{h}_i$ is deterministic and represents the power loss over a propagation path of length $d_i$ and is given by~\cite{Steve_pointing_error,MyTCOM}
\begin{IEEEeqnarray}{lll}\label{Eq:pathloss}
\bar{h}_i=10^{-\kappa d_i/10},
\end{IEEEeqnarray}
where $\kappa$ is the weather-dependent attenuation factor of the FSO links.

\subsubsection{Geometric Loss}
 $\hat{h}_{i}$ is also deterministic and is caused by the divergence  of the optical beam between transmitter and PD. It is modeled as~\cite{Steve_pointing_error,MyTCOM}
\begin{IEEEeqnarray}{lll}\label{Eq:geometric_loss}
\hat{h}_i=\left[\mathrm{erf}\left(\dfrac{\sqrt{\pi}r}{\sqrt{2}\phi d_{i}}\right)\right]^2,
\end{IEEEeqnarray}
where $\mathrm{erf}(\cdot)$ is the error function and $\phi$ is the divergence angle of the beam.

\subsubsection{Atmospheric Turbulence}
We use the G-G model for the atmospheric turbulence $\tilde{h}_i$, since the G-G distribution is able to accurately model a wide range of weak to strong turbulence conditions~\cite{Steve_pointing_error}. In particular, the  probability density function (PDF) of $\tilde{h}_i$ is given by~\cite{GG_George}
\begin{IEEEeqnarray}{lll}\label{Eq:Turbulence}
f_{\tilde{h}_i}(\tilde{h}_i)=\dfrac{2(\alpha\beta)^{(\alpha+\beta)/2}}{\Gamma(\alpha)\Gamma(\beta)}\tilde{h}_i^{(\alpha+\beta)/2-1}K_{\alpha-\beta}\left(2\sqrt{\alpha\beta\tilde{h}_i}\right),\quad
\end{IEEEeqnarray}
where  $\Gamma(\cdot)$ is the Gamma function, $K_{\alpha-\beta}(\cdot)$ is the modified Bessel function of the second kind, and $\alpha$ and $\beta$ depend
on physical parameters such as the wavelength $\lambda^{\mathrm{fso}}$ and the weather-dependent index of refraction structure parameter $C_n^2$  \cite{MyTCOM}. The cumulative  distribution function (CDF) of $\tilde{h}_i$ is given by
\begin{IEEEeqnarray}{lll}\label{Eq:Turbulence_CDF}
F_{\tilde{h}_i}(\tilde{h}_i)=\dfrac{1}{\Gamma(\alpha)\Gamma(\beta)}G_{1,3}^{2,1}\left[\alpha\beta\tilde{h}_i\Big|^{1}_{\alpha,\beta,0}\right],
\end{IEEEeqnarray}
where $G_{a,b}^{c,d}(\cdot)$ is the Meijer's G-function \cite{TableIntegSerie8}.
 
\section{NOMA for FSO Backhauling}
In this section, we first present the electrical signal-to-interference-plus-noise ratio (SINR) at the CU. Subsequently, we derive the optimal dynamic-order decoding strategy.

\subsection{Electrical SINR at the CU}
Since the capacity of a point-to-point IM/DD FSO link is not known in general, we consider the following achievable rate \cite{FSO_Cap}
\begin{IEEEeqnarray}{lll}\label{Eq:Rate}
R(\Gamma_i)=\dfrac{1}{2}\log_2\left(1+\dfrac{e}{2\pi}\Gamma_i\right)\quad\text{bits/symbol},
\end{IEEEeqnarray}
where $\Gamma_i$ is the SINR of BS$_i$ at the CU. We note that $\Gamma_i$ depends on the adopted decoding order. For future reference, let $\mathbf{o}=(i,i')$ denote the  decoding order if BS$_i$ is decoded first and BS$_{i'}$ is decoded second. Moreover, assuming that BS$_i$ is decoded second, $\Gamma_i$  depends on  whether the SIC is successful or erroneous. To formally take the effect of the reliability of SIC into account, we define binary variable $s$ which is equal to one if the decoding of the first signal is successful; otherwise, it is equal zero. Using these definitions, under the perfect, imperfect, and worst-case SIC assumptions \cite{SIC_NOMA,Dynamic_SIC_NOMA,NOMA_Survey_Robert}, the SINRs of the BSs are obtained as
\begin{IEEEeqnarray}{lll}\label{Eq:SINRNOMA}
\Gamma_1 = \begin{cases}
\dfrac{\gamma_1}{\gamma_2+1},\,\,&\mathrm{if} \,\,\mathbf{o}=(1,2) \\
\gamma_1,\,\,
&\mathrm{if} \,\,\mathbf{o}=(2,1)\,\,\text{for perfect SIC} \\
\dfrac{\gamma_1}{(1-s)\gamma_2+1},\,\,&\mathrm{if} \,\,\mathbf{o}=(2,1)\,\,\text{for imperfect SIC} \\
s\gamma_1,\,\,
&\mathrm{if} \,\,\mathbf{o}=(2,1)\,\,\text{for worst-case SIC,}
\end{cases}
\IEEEyesnumber\IEEEyessubnumber \\
\Gamma_2 = \begin{cases}
\dfrac{\gamma_2}{\gamma_1+1},\,\,&\mathrm{if} \,\,\mathbf{o}=(2,1) \\
\gamma_2,\,\,
&\mathrm{if} \,\,\mathbf{o}=(1,2)\,\,\text{for perfect SIC}\\
\dfrac{\gamma_2}{(1-s)\gamma_1+1},\,\,&\mathrm{if} \,\,\mathbf{o}=(1,2)\,\,\text{for imperfect SIC} \\
s\gamma_2,\,\,
&\mathrm{if} \,\,\mathbf{o}=(1,2)\,\,\text{for worst-case SIC,}
\end{cases} \quad \IEEEyesnumber\IEEEyessubnumber
\end{IEEEeqnarray}
where $\gamma_1 = \frac{P_1^2|h_1|^2}{\delta^2_n}$ and $\gamma_2 = \frac{P_2^2|h_2|^2}{\delta^2_n}$.

\subsection{Dynamic-Order Decoding} 

We assume that the BSs transmit with a fixed rate denoted by $R_i^{\mathrm{thr}},\,\, i=1,2,$  (in bits/symbol) regardless of the channel conditions. Therefore, CSI is needed only at the CU. In this case, outage probability is commonly adopted as performance criterion \cite{Steve_pointing_error}. Here, the decoding order is chosen according to the CSI at the CU and the  QoS requirements of the BSs.
In the following proposition, we provide the optimal decoding order which minimizes the outage probabilities of the BSs, $P^{\mathrm{out}}_i,\,i=1,2,$ under the perfect, imperfect, and worst-case SIC assumptions, respectively. 

\begin{prop}\label{Prop:Opt_order}
The optimal policy for the SIC decoding order which jointly minimizes $P^{\mathrm{out}}_1$ and $P^{\mathrm{out}}_2$ is identical under the perfect, imperfect, and worst-case SIC assumptions and is given by 
\begin{IEEEeqnarray}{lll}\label{Eq:AdaptiveOrder}
\mathbf{o}=
\begin{cases}
(1,2),\,\, &\mathrm{if}\,\, \hat{\gamma}_{1}\geq\gamma_1^{\mathrm{thr}} \,\,\&\,\,\hat{\gamma}_{2}<\gamma_2^{\mathrm{thr}} \\
(2,1),\,\, &\mathrm{if}\,\, \hat{\gamma}_{1}<\gamma_1^{\mathrm{thr}} \,\,\&\,\,\hat{\gamma}_{2}\geq\gamma_2^{\mathrm{thr}} \\
(1,2) \,\,\text{or}\,\, (2,1),\, &\mathrm{otherwise,} 
\end{cases}\quad
\end{IEEEeqnarray}
 where $\hat{\gamma}_1=\frac{\gamma_1}{\gamma_2+1}$, $\hat{\gamma}_2=\frac{\gamma_2}{\gamma_1+1}$, and $\gamma_i^{\mathrm{thr}} = \frac{e}{2\pi}\big(2^{2R_i^{\mathrm{thr}}}-1\big)$. 
\end{prop}
\begin{IEEEproof}
Please refer to Appendix~\ref{App:Prop_Order}.
\end{IEEEproof}

The main idea behind the optimal decoding strategy in Proposition~\ref{Prop:Opt_order} is to choose the decoding order in which the signal of the BS with better channel conditions and lower QoS requirements is decoded first. 
To better understand Proposition~\ref{Prop:Opt_order}, we distinguish the following cases. Regardless of the adopted decoding order, if $\hat{\gamma}_{1}\geq\gamma_1^{\mathrm{thr}}$ and $\hat{\gamma}_{2}\geq\gamma_2^{\mathrm{thr}}$ hold, both BSs are not in outage and if $\hat{\gamma}_{1} <\gamma_1^{\mathrm{thr}}$ and $\hat{\gamma}_{2}<\gamma_2^{\mathrm{thr}}$ hold, both BSs are  in outage. For the other two possibilities $\hat{\gamma}_{1}\geq\gamma_1^{\mathrm{thr}}$ and $\hat{\gamma}_{2}<\gamma_2^{\mathrm{thr}}$ or $\hat{\gamma}_{1}<\gamma_1^{\mathrm{thr}}$ and $\hat{\gamma}_{2}\geq\gamma_2^{\mathrm{thr}}$, the optimal decoding orders are $(1,2)$ and $(2,1)$, respectively, where SIC is always successful, i.e., $s=1$.
Therefore, for the dynamic-order decoding strategy in (\ref{Eq:AdaptiveOrder}), the outage probabilities of the BSs do not depend on whether perfect, imperfect, or worst-case SIC is assumed. 
 
\begin{figure}
\centering
\scalebox{0.75}{
\pstool[width=1.3\linewidth]{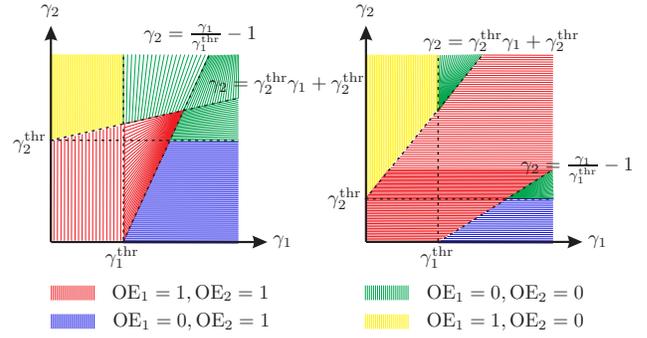}{
\psfrag{g2}[c][c][1]{$\gamma_2$}
\psfrag{g1}[c][c][1]{$\gamma_1$}
\psfrag{Oe1}[l][c][1]{$\mathrm{OE}_1=1,\mathrm{OE}_2=1$}
\psfrag{Oe2}[l][c][1]{$\mathrm{OE}_1=0,\mathrm{OE}_2=1$}
\psfrag{Oe3}[l][c][1]{$\mathrm{OE}_1=0,\mathrm{OE}_2=0$}
\psfrag{Oe4}[l][c][1]{$\mathrm{OE}_1=1,\mathrm{OE}_2=0$}
\psfrag{gt1}[c][c][1]{$\gamma_1^{\mathrm{thr}}$}
\psfrag{gt2}[c][c][1]{$\gamma_2^{\mathrm{thr}}$}
\psfrag{lin1}[c][l][1]{$\gamma_2=\frac{\gamma_1}{\gamma_1^{\mathrm{thr}}}-1$}
\psfrag{lin2}[c][l][1]{$\gamma_2=\gamma_2^{\mathrm{thr}}\gamma_1+\gamma_2^{\mathrm{thr}}$}
}}
\caption{Illustration of the outage events for BS$_i$ (OE$_i$), $i=1,2$, for the optimal decoding order for $\gamma_1^{\mathrm{thr}}\gamma_2^{\mathrm{thr}}<1$ (the left-hand side figure) and $\gamma_1^{\mathrm{thr}}\gamma_2^{\mathrm{thr}}\geq 1$ (the right-hand side figure).}
\label{FigOE_adaptive}\vspace{-0.3cm}
\end{figure}

\section{Performance Analysis under G-G Turbulence}
 
 In this section, we first analyze the outage probability for  arbitrary SNRs and then simplify the results for the high SNRs.

\subsection{General Case} 

We emphasize that the outage probabilities of the BSs for the optimal decoding order scheme in Proposition~\ref{Prop:Opt_order} do not depend on the adopted SIC assumption, cf. Section~III. In particular,  the outage probability of the BSs are given by
\begin{IEEEeqnarray}{lll}\label{Eq:outageNOMA_adaptive}
P^{\mathrm{out}}_1&=\Pr(\Gamma_1<\gamma_1^{\mathrm{thr}})  \IEEEyesnumber\IEEEyessubnumber\\ &= \mathrm{Pr}\left(\hat{\gamma}_{1}<\gamma_1^{\mathrm{thr}},\hat{\gamma}_{2}<\gamma_2^{\mathrm{thr}}\right) +  \mathrm{Pr}\left(\hat{\gamma}_{2}\geq\gamma_2^{\mathrm{thr}},\gamma_{1}<\gamma_1^{\mathrm{thr}}\right) \hspace{-5mm} \nonumber \\
P^{\mathrm{out}}_2&=\Pr(\Gamma_2<\gamma_2^{\mathrm{thr}}) \IEEEyessubnumber\\ &=\mathrm{Pr}\left(\hat{\gamma}_{1}<\gamma_1^{\mathrm{thr}},\hat{\gamma}_{2}<\gamma_2^{\mathrm{thr}}\right) 
+  \mathrm{Pr}\left(\hat{\gamma}_{1}\geq\gamma_1^{\mathrm{thr}},\gamma_{2}<\gamma_2^{\mathrm{thr}}\right). \hspace{-5mm} \quad\nonumber 
\end{IEEEeqnarray}
 Fig.~\ref{FigOE_adaptive} schematically illustrates the outage events in the plane of $\gamma_1$-$\gamma_2$, for the cases of $\gamma_1^{\mathrm{thr}}\gamma_2^{\mathrm{thr}}<1$ and $\gamma_1^{\mathrm{thr}}\gamma_2^{\mathrm{thr}}\geq1$, where lines $\gamma_2=\frac{\gamma_1}{\gamma_1^{\mathrm{thr}}}-1$ and $\gamma_2=\gamma_2^{\mathrm{thr}}\gamma_1+\gamma_2^{\mathrm{thr}}$  intersect and do not intersect, respectively. Here,  $\mathrm{OE}_i=1$ and $\mathrm{OE}_i=0$ indicate that an outage event occurs and does  not occur for BS$_i,\,i=1,2$, respectively. In particular, from Fig.~\ref{FigOE_adaptive},  an outage occurs for BS$_1$ in the red and blue regions, i.e., $P^{\mathrm{out}}_1=\mathrm{Pr}\left(\mathrm{OE}_1=1,\mathrm{OE}_2=1\right)+\mathrm{Pr}\left(\mathrm{OE}_2=0,\mathrm{OE}_1=1\right)$.  Similarly, an outage happens for BS$_2$ in the red and yellow regions, i.e., $P^{\mathrm{out}}_2=\mathrm{Pr}\left(\mathrm{OE}_1=1,\mathrm{OE}_2=1\right)+\mathrm{Pr}\left(\mathrm{OE}_1=0,\mathrm{OE}_2=1\right)$.  
 
We first present the following lemma, which facilitates a numerical evaluation of the outage probabilities in  (\ref{Eq:outageNOMA_adaptive}).

\begin{lem}\label{Lemm_pdfY}
Let $X$ and $Y$ be independent G-G random variables (RVs) and $a,b,c$, and $d$ be constants. The following identities hold
\begin{IEEEeqnarray}{lll}\label{Eq:pdfY}
f_1(X,Y,a,b,c,d)\triangleq \mathrm{Pr}\left(\dfrac{aX^2}{bY^2+1}\geq c,bY^2< d\right) \nonumber\\
\qquad\quad=\int_{0}^{\sqrt{d/b}}\left[1-F_{X}\left(\sqrt{c(by^2+1)/a}\right)\right]f_{Y}(y)\mathrm{d}y, \qquad\\
f_2(X,Y,a,b,c,d)\triangleq\mathrm{Pr}\left(\dfrac{aX^2}{bY^2+1}\leq c, \dfrac{bY^2}{aX^2+1} \leq d\right) \nonumber\\
\qquad\quad=\int_0^{\zeta}F_{X}\left(\sqrt{c(by^2+1)/a}\right)f_{Y}(y)\mathrm{d}y \nonumber \\
\qquad\quad-\int_{\sqrt{d/b}}^{\zeta}F_{X}\left(\sqrt{\Big(\frac{b}{d}y^2-1\Big)/a}\right)f_{Y}(y)\mathrm{d}y,\quad
\end{IEEEeqnarray}
 where $\zeta$ is a constant given by
\begin{IEEEeqnarray}{lll} 
\zeta = \begin{cases}
\sqrt{\frac{d(1+c)}{b(1-cd)}},\,\,&\mathrm{if}\,\,cd<1 \\
\infty, \,\,&\mathrm{otherwise}.
\end{cases}
\end{IEEEeqnarray}
Moreover, $F_{X}(\cdot)$ and $f_{Y}(\cdot)$ are the CDF and PDF of G-G RVs $X$ and $Y$ and are given in (\ref{Eq:Turbulence_CDF}) and (\ref{Eq:Turbulence}), respectively. 
\end{lem}
\begin{IEEEproof}
Please refer to Appendix~\ref{App:Lemm}. 
\end{IEEEproof}

We note that $\gamma_1$ and $\gamma_2$ are two independent RVs which can be rewritten as $e_1\tilde{h}_1^2$ and $e_2\tilde{h}_2^2$, respectively, where $e_i=\frac{P_i^2\bar{h}_i^2\hat{h}_i^2}{\delta_n^2}, \,\,i=1,2$, is a constant and $\tilde{h}_1$ and $\tilde{h}_2$ are independent G-G RVs. Therefore, using Lemma~\ref{Lemm_pdfY}, the outage probability of BS$_i$  given in (\ref{Eq:outageNOMA_adaptive}) is obtained as follows
\begin{IEEEeqnarray}{lll} 
P_1^{\mathrm{out}} =& f_2\left(\tilde{h}_1,\tilde{h}_2,e_1,e_2,\gamma_1^{\mathrm{thr}},\gamma_2^{\mathrm{thr}}\right)\nonumber\\
&+f_1\left(\tilde{h}_2,\tilde{h}_1,e_2,e_1,\gamma_2^{\mathrm{thr}},\gamma_1^{\mathrm{thr}}\right) \qquad \IEEEyesnumber\IEEEyessubnumber \\
P_2^{\mathrm{out}} = &f_2\left(\tilde{h}_1,\tilde{h}_2,e_1,e_2,\gamma_1^{\mathrm{thr}},\gamma_2^{\mathrm{thr}}\right)  \nonumber \\
&+ f_1\left(\tilde{h}_1,\tilde{h}_2,e_1,e_2,\gamma_1^{\mathrm{thr}},\gamma_2^{\mathrm{thr}}\right). \qquad\,\IEEEyessubnumber
\end{IEEEeqnarray}
Using the above expressions, we are able to numerically compute the outage probabilities and verify our simulation~results.

\subsection{High SNR Regime}
In this subsection, we provide an outage analysis for the high SNR regime, i.e., when $\gamma\to\infty$ where $\gamma=\frac{P_1^2}{\delta_n^2}=\frac{P_2^2}{\delta_n^2}$. 
\begin{corol}\label{crol:Asym}
The outage probability of the BSs in the high SNR regime, i.e., when  $\gamma\to\infty$, for the proposed optimal dynamic NOMA scheme is given by
\begin{IEEEeqnarray}{lll}\label{Eq:outageNOMA_adaptive_Asym}
\lim_{\gamma\to \infty}P^{\mathrm{out}}_1=\nonumber\\
\quad \begin{cases}
F_1\left(\alpha\beta\sqrt{\frac{\gamma_1^{\mathrm{thr}}}{c_1\gamma}}\right), &\mathrm{if}\, \gamma_1^{\mathrm{thr}}\gamma_2^{\mathrm{thr}}<1 \\
F_2\left(\sqrt{c\gamma_1^{\mathrm{thr}}}\right)-F_2\left(\sqrt{\frac{c}{\gamma_2^{\mathrm{thr}}}}\right), &\mathrm{if}\, \gamma_1^{\mathrm{thr}}\gamma_2^{\mathrm{thr}}\geq1,
\end{cases}\IEEEyesnumber\IEEEyessubnumber\\
\lim_{\gamma\to \infty}P^{\mathrm{out}}_2=\nonumber\\
\quad \begin{cases}
F_1\left(\alpha\beta\sqrt{\frac{\gamma_2^{\mathrm{thr}}}{c_2\gamma}}\right),\ &\mathrm{if}\, \gamma_1^{\mathrm{thr}}\gamma_2^{\mathrm{thr}}<1 \\
F_2\left(\sqrt{c\gamma_1^{\mathrm{thr}}}\right)-F_2\left(\sqrt{\frac{c}{\gamma_2^{\mathrm{thr}}}}\right), &\mathrm{if}\, \gamma_1^{\mathrm{thr}}\gamma_2^{\mathrm{thr}}\geq1,\IEEEyessubnumber
\end{cases}
\end{IEEEeqnarray}
respectively, where $c=\frac{c_2}{c_1}$ and $c_i=\bar{h}_i^2\hat{h}_i^2,\,\,i=1,2$. Moreover, $F_1(x)$ and  $F_2(x)$  are given by
\begin{IEEEeqnarray}{lll}
F_1(x)=\frac{1}{\Gamma(\alpha)\Gamma(\beta)}\Bigg[\frac{\Gamma(\beta-\alpha)}{\alpha} x^{\alpha}
+\frac{\Gamma(\alpha-\beta)}{\beta}x^{\beta}\Bigg]\label{Eq:F1}\\
F_2(x)=\frac{1}{\Gamma(\alpha)^2\Gamma(\beta)^2}G_{3,3}^{2,3}\left[x\Big\vert^{1-\alpha,1-\beta,1}_{\alpha,\beta,0}\right].\label{Eq:F2}
\end{IEEEeqnarray}
\end{corol}
\begin{IEEEproof}
Please refer to Appendix~\ref{App:Asym}.
\end{IEEEproof}

Corollary~1 reveals that the proposed optimal dynamic NOMA scheme exhibits two different asymptotic behaviors depending on the  QoS requirements. \textit{i)} If $\gamma_1^{\mathrm{thr}}\gamma_2^{\mathrm{thr}}<1$ holds, the outage probabilities do not exhibit an outage floor. We note that one can further simplify the asymptotic expression in (\ref{Eq:F1}) by neglecting the term associated with $x^{\alpha}$ if $\alpha>\beta$ or the term associated with $x^{\beta}$ if $\alpha<\beta$. Therefore, when $\gamma_1^{\mathrm{thr}}\gamma_2^{\mathrm{thr}}<1$,
the diversity order of the outage probabilities of the BSs is $\frac{1}{2}\min\{\alpha,\beta\}$. \textit{ii)} If $\gamma_1^{\mathrm{thr}}\gamma_2^{\mathrm{thr}}\geq1$ holds, the outage probabilities have an outage floor. Note that the value of the outage floor is the same for both BSs and depends on the ratio of the average channel gains, $c$, as well as the required QoS, $\gamma^{\mathrm{thr}}_i$.

\begin{remk}
Note that in NOMA, in general, the signal of the BS which is decoded first always suffers from interference from the other BS which may cause an outage floor in high SNR. However, Corollary 1 reveals that under certain conditions, i.e., $\gamma_1^{\mathrm{thr}}\gamma_2^{\mathrm{thr}}<1$,
an outage floor does not occur since in high SNR, the signal of at least \textit{one} of the BSs can be correctly decoded. In other words, in high SNR, for all channel realizations where the signal of BS$_1$ cannot be decoded first, the signal of BS$_2$ very likely can be correctly decoded first instead, and vice versa.
\end{remk}

\section{Performance Evaluation}
In this section, we first present our baseline and benchmark schemes and subsequently we provide simulation results to evaluate the performance of the proposed NOMA scheme.

\subsection{Baseline and Benchmark Schemes}
We consider the following benchmark/baseline schemes:

\textit{OMA:} Here,  BS$_1$ is active half of the available time and remains silent when BS$_2$ is active in the other half of the available time. In order to have a fair comparison with NOMA, we scale the transmission rates of the BSs, such that the average data rates of both schemes are equal, i.e., with OMA the transmission rate of BS$_i$ is $2R^{\mathrm{thr}}_i$.

\textit{Fixed NOMA:} We consider a simple NOMA scheme which employs fixed-order decoding. Here, we assume the signal of BS$_1$, which is closer to the CU, is decoded first and the signal of BS$_2$ is decoded second.

\textit{Dynamic NOMA scheme in \cite{arXiv_NOMA_FSO}:} We consider the NOMA scheme in \cite{arXiv_NOMA_FSO} which performs dynamic SIC ordering based on the instantaneous received powers from the BSs, i.e., the signal of the BS with higher instantaneous received power is decoded first. 

\textit{Performance upper bound:}  To have a lower bound on the outage probability, we assume that both BSs transmit concurrently; however, their signals can be decoded interference free at the CU. Note that this provides an upper bound on the performance of any NOMA and OMA scheme. 

For the baseline NOMA schemes, we present the results under the imperfect SIC assumption.
\begin{table}
\label{Table:Parameter}
\caption{Simulation Parameters~\cite{MyTCOM,FSO_Vahid}.\vspace{-0.2cm}} 
\begin{center}
\scalebox{0.72}
{
\begin{tabular}{|| c | c  | c ||}
  \hline
 Symbol & Definition & Value \\ \hline \hline
  $\delta^2_n$ & Noise variance at the CU & $10^{-14}$ $\mathrm{A}^2$ \\ \hline 
       $R$ & Responsivity of PD at the CU & $0.5\frac{1}{\mathrm{V}}$   \\ \hline  
       $\phi$ & Laser divergence angle  & $2$ mrad  \\ \hline  
       $r$ & Aperture radius  & $10$ cm  \\ \hline      
              $(\alpha,\beta)$ & Parameters of G-G fading  & $(2.23,1.54)$  \\ \hline  
\end{tabular}
}
\end{center}
\vspace{-0.7cm}
\end{table}
\subsection{Simulation Results}
 Unless stated otherwise, the values of the FSO link parameters used in our simulations are given in Table I. 

\begin{figure}
  \centering
\resizebox{0.9\linewidth}{!}{\psfragfig{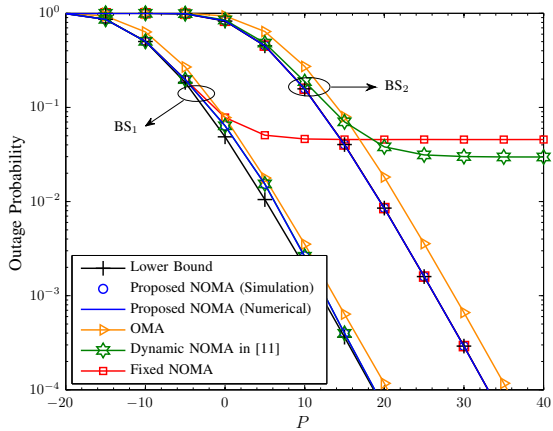}} 
\vspace{-5mm}
\caption{Outage probability of BS$_1$ and BS$_2$ vs. optical transmit power  $P_1=P_2=P$ for  $(d_1,d_2)=(1000,2000)$ m, $(R_1^{\mathrm{thr}},R_2^{\mathrm{thr}})=(0.1,0.5)$ bits/symbol, and $\kappa=4.2\times10^{-3}$ (haze).} \vspace{-2mm}
\label{Fig:Outage_k4}
\end{figure}

 In Fig.~\ref{Fig:Outage_k4}, we plot the outage probability of the BSs vs. optical transmit power $P_1=P_2=P$ in dBm for $\kappa=4.2\times 10^{-3}$ (haze), $(d_1,d_2)=(1000,2000)$ m, and $(R_1^{\mathrm{thr}},R_2^{\mathrm{thr}})=(0.1,0.5)$ bits/symbol. We observe from Fig.~\ref{Fig:Outage_k4} that for both baseline NOMA schemes, the outage probability of one of the BSs has an outage floor whereas for the proposed dynamic NOMA, an outage floor does not occur for either BS. This is due to the fact that the QoS requirements of the BSs lead to $\gamma^{\mathrm{thr}}_1\gamma^{\mathrm{thr}}_2=0.7945<1$, cf. Corollary~1.  In addition, for OMA each BS is active half of the time, and thus, we expect a maximum SNR loss of $3$~dB with respect to the lower bound which is consistent with the results shown in Fig.~\ref{Fig:Outage_k4}. Note that since we plot the outage probability vs. the \textit{optical transmit power}, the slope of the outage curves for the proposed NOMA scheme is $\beta = 1.54$ which is twice the diversity order obtained in Corollary~\ref{crol:Asym} for the electrical power.
Moreover, for the proposed dynamic NOMA, the outage probability of BS$_1$ is less than the outage probability of BS$_2$ which is due to the fact that both the distance from the CU and the required QoS are less for BS$_1$ than for BS$_2$.
Furthermore, the performance of the proposed scheme is close to the lower bound especially for high SNR, which reveals the effectiveness of the proposed NOMA scheme and its relative resilience to multiuser interference. Finally, we observe a perfect match between simulation and numerical results, which validates our derivations. 

\begin{figure}
  \centering
\resizebox{0.9\linewidth}{!}{\psfragfig{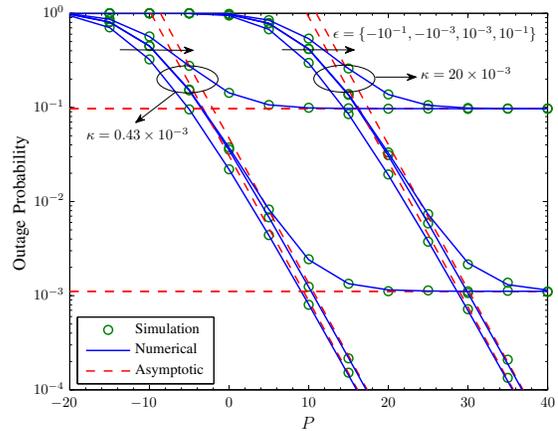}} 
\vspace{-5mm}
\caption{Outage probability, $P^{\mathrm{out}}_1=P^{\mathrm{out}}_2$, vs. optical transmit power $P_1=P_2=P$ for $d_1=d_2=1000$~m and
 $R^{\mathrm{thr}}_1=R^{\mathrm{thr}}_2=R^{\mathrm{thr}}_{\mathrm{crt}}+\epsilon$. The value of $\epsilon$ increases along the direction of the arrows. } \vspace{-2mm}
\label{Fig:Outage_k20}
\end{figure}

Next, we study the effects of different QoS requirements and FSO weather conditions. For clarity of presentation, we assume $P_1=P_2=P$~dBm, $d_1=d_2=1000$~m, and
 $R^{\mathrm{thr}}_1=R^{\mathrm{thr}}_2$  which leads to $P^{\mathrm{out}}_1=P^{\mathrm{out}}_2$. Fig.~\ref{Fig:Outage_k20} shows the outage probability and the asymptotic outage probability  given in Corollary~1 for the proposed dynamic NOMA scheme vs. the optical transmit power $P$ for two different weather conditions, i.e., $\kappa=0.43\times 10^{-3}$ (clear air) and $\kappa=20\times 10^{-3}$ (light fog). Here, we assume 
 $R^{\mathrm{thr}}_i=R^{\mathrm{thr}}_{\mathrm{crt}}+\epsilon$, where $R^{\mathrm{thr}}_{\mathrm{crt}}$ is the critical rate given by $R^{\mathrm{thr}}_{\mathrm{crt}}=\frac{1}{2}\log_2(1+\frac{e}{2\pi})$ for which $\gamma^{\mathrm{thr}}_1=\gamma^{\mathrm{thr}}_2=1$ holds. We can observe from Fig.~\ref{Fig:Outage_k20} that as expected, for $\epsilon<0$, i.e., $\gamma^{\mathrm{thr}}_1\gamma^{\mathrm{thr}}_2<1$, there is no outage floor for  high SNR and for $\epsilon\geq0$, i.e., $\gamma^{\mathrm{thr}}_1\gamma^{\mathrm{thr}}_2\geq1$, an outage floor exists. Moreover, this figure shows that the more stringent the QoS requirements are, i.e., the higher $R^{\mathrm{thr}}_i$, the higher the outage probability becomes. Moreover, as expected, as the weather conditions deteriorate, i.e., for larger values of $\kappa$, the outage probability increases which manifests itself in an SNR loss. Finally, all outage probability curves approach the asymptotic results derived in Corollary~1. Again, we observe a perfect agreement between the numerical and simulation results.
\section{Conclusion}
In this paper, we considered NOMA for backhauling of two BSs to a CU via FSO links. We derived an optimal dynamic NOMA scheme which jointly minimizes the BSs' outage probabilities. Moreover, we analyzed the outage probability for G-G turbulence and derived closed-form expressions in the high SNR regime. Our simulation results validated the analytical derivations and revealed that the proposed dynamic NOMA scheme can achieve a considerable performance gain compared to OMA as well as fixed and dynamic NOMA schemes from the literature.

\appendices

\section{} \label{App:Prop_Order}
We distinguish the following four mutually exclusive cases depending on the CSI and the QoS requirements of the BSs.

\noindent
\textit{Case 1:} $\hat{\gamma}_{1}\geq\gamma_1^{\mathrm{thr}} \,\,\&\,\,\hat{\gamma}_{2}<\gamma_2^{\mathrm{thr}}$:  In this case, for decoding order $(1,2)$, we obtain $s=1$ whereas for decoding order $(2,1)$, we obtain $s=0$. 

\textit{i)} Under the perfect SIC assumption, we obtain $\Gamma_1=\hat{\gamma}_{1}\geq\gamma_1^{\mathrm{thr}}$ and $\Gamma_1=\gamma_{1}\geq\gamma_1^{\mathrm{thr}}$ for orders $(1,2)$ and $(2,1)$, respectively, i.e., BS$_1$ is not in outage regardless of the decoding order. On the other hand,  we obtain  $\Gamma_2=\hat{\gamma}_{2}<\gamma_2^{\mathrm{thr}}$ for order $(2,1)$ which leads to an outage for BS$_2$, whereas for decoding order $(1,2)$, we obtain $\Gamma_2=\gamma_2$ where BS$_2$ may or may not be in outage. Therefore, order $(1,2)$ is optimal. 

\textit{ii)} Under the imperfect SIC assumption, we obtain $\Gamma_1=\hat{\gamma}_{1}\geq\gamma_1^{\mathrm{thr}}$ for both possible decoding orders, i.e., regardless of the decoding order, BS$_1$ is not in outage. On the other hand, for decoding order $(2,1)$, we obtain $\Gamma_2=\hat{\gamma}_2<\gamma_2^{\mathrm{thr}}$ which leads to an outage for BS$_2$ whereas for decoding order $(1,2)$, we obtain $\Gamma_2=\gamma_2$ where BS$_2$ may or may not be in outage. Therefore, the optimal decoding order is $(1,2)$.

\textit{iii)} Under the worst-case SIC assumption, for order $(2,1)$, we obtain $\Gamma_1=0<\gamma_1^{\mathrm{thr}}$ and $\Gamma_2=\hat{\gamma}_{2}<\gamma_2^{\mathrm{thr}}$ which leads to an outage for both BSs. On the other hand, for order $(1,2)$, we obtain $\Gamma_1=\hat{\gamma}_{1}\geq\gamma_1^{\mathrm{thr}}$ which means BS$_1$ is not in outage. Therefore, order $(1,2)$ is optimal.

\noindent
\textit{Case 2:} $\hat{\gamma}_{1}<\gamma_1^{\mathrm{thr}} \,\,\&\,\,\hat{\gamma}_{2}\geq\gamma_2^{\mathrm{thr}}$: In this case, the optimal order is $(2,1)$. The proof is similar to that for Case 1 after switching the roles of BS$_1$ and BS$_2$.

\noindent
\textit{Case 3:} $\hat{\gamma}_{1}\geq\gamma_1^{\mathrm{thr}} \,\,\&\,\,\hat{\gamma}_{2}\geq\gamma_2^{\mathrm{thr}}$:
In this case, we obtain $s=1$ for both  decoding orders. This leads to $\Gamma_i\geq\gamma_i^{\mathrm{thr}},\,\,i=1,2$, for both  decoding orders under the perfect, imperfect, and worst-case SIC assumptions. In other words,  regardless of the decoding order and the SIC assumption, both BSs will not be in outage. 

\noindent
\textit{Case 4:} $\hat{\gamma}_{1}<\gamma_1^{\mathrm{thr}} \,\,\&\,\,\hat{\gamma}_{2}<\gamma_2^{\mathrm{thr}}$:
In this case, we obtain $s=0$ for both  decoding orders. This leads to $\Gamma_i<\gamma_i^{\mathrm{thr}},\,\,i=1,2$, for both  decoding orders  under the perfect, imperfect, and worst-case SIC assumptions. In other words,  regardless of the decoding order and the SIC assumption, both BSs will be in outage. 

The above decoding order strategy is concisely given in (\ref{Eq:AdaptiveOrder}) which completes the proof.

\section{}\label{App:Lemm}
Due to space constraints, we provide the proof only for the more involved case of $f_2(X,Y,a,b,c,d)$ and skip the proof for  $f_1(X,Y,a,b,c,d)$ which is similar. We note that for $\gamma_1=aX^2$ and $\gamma_2=bY^2$, $f_2(X,Y,a,b,c,d)$ corresponds to the region with red color in Fig.~\ref{FigOE_adaptive}. Therefore, we obtain
\begin{IEEEeqnarray}{lll} 
f_2(X,Y,a,b,c,d)\triangleq\mathrm{Pr}\left(\dfrac{aX^2}{bY^2+1}\leq c, \dfrac{bY^2}{aX^2+1}\leq d\right)\nonumber \\
\qquad\quad=\mathrm{Pr}\left(X\leq \sqrt{\dfrac{c(bY^2+1)}{a}}, X\geq \sqrt{\dfrac{\frac{b}{d}Y^2-1}{a}}\right)\nonumber \\
\qquad\quad=\int_0^{\zeta} \mathrm{Pr}\left(X\leq \sqrt{\dfrac{c(by^2+1)}{a}}\big|Y=y\right) f_{Y}(y)\mathrm{d}y\nonumber \\
\qquad\quad-\int_{\sqrt{d/b}}^{\zeta} \mathrm{Pr}\left(X\leq \sqrt{\dfrac{\frac{b}{d}y^2-1}{a}}\big|Y=y\right) f_{Y}(y)\mathrm{d}y\nonumber \\
\qquad\quad= \int_0^{\zeta}F_{X}\left(\sqrt{c(by^2+1)/a}\right)f_{Y}(y)\mathrm{d}y \nonumber \\
\qquad\quad-\int_{\sqrt{d/b}}^{\zeta}F_{X}\left(\sqrt{\Big(\frac{b}{d}y^2-1\Big)/a}\right)f_{Y}(y)\mathrm{d}y,
\end{IEEEeqnarray}
where $\zeta$ is the value of $Y$ for the intersection of $aX^2/(bY^2+1)=c$ and $bY^2/(aX^2+1)=d$. Here, if $cd<1$ holds, the aforementioned functions intersect at a unique point resulting in $\zeta=\frac{d(1+c)}{b(1-cd)}$, otherwise, the two functions do not intersect and we have $\zeta = \infty$. This completes the proof.

\section{} \label{App:Asym}
Without loss of generality, we  analyze only $P^{\mathrm{out}}_1$, since $P^{\mathrm{out}}_2$ can be obtained straightforwardly after switching the roles of BS$_1$ and BS$_2$.
Using $\hat{\gamma}_1 \to \frac{\gamma_1}{\gamma_2}$ and $\hat{\gamma}_2 \to \frac{\gamma_2}{\gamma_1}$ which hold for high SNR, (\ref{Eq:outageNOMA_adaptive}a) can be written as $P^{\mathrm{out}}_1=A+B$ where $A$ and $B$ are as follows
\begin{IEEEeqnarray}{lll}
A= \mathrm{Pr}\Big(\frac{\gamma_{1}}{\gamma_{2}}<\gamma_1^{\mathrm{thr}},\frac{\gamma_{2}}{\gamma_{1}}<{\gamma_2^{\mathrm{thr}}}\Big) \nonumber \\
B= \mathrm{Pr}\Big(\gamma_{1}<\gamma_1^{\mathrm{thr}},\frac{\gamma_{2}}{\gamma_{1}}\geq\gamma_2^{\mathrm{thr}}\Big).
\end{IEEEeqnarray}
Now, we distinguish the following two cases:

\noindent
\textit{Case~1:} If $\gamma_1^{\mathrm{thr}}\gamma_2^{\mathrm{thr}}<1$ holds, $A$ becomes zero and $B$ approaches $\mathrm{Pr}\big(\gamma_{1}<{\gamma_1^{\mathrm{thr}}}\big)$. Moreover, after some manipulations  $B=F_{\tilde{h}_1}\Big(\sqrt{\frac{\gamma_1^{\mathrm{thr}}}{c_1\gamma}}\Big)$, where $F_{\tilde{h}_1}(\cdot)$ is the CDF of G-G RV $\tilde{h}_1$  given in (\ref{Eq:Turbulence_CDF}). Then, we approximate the CDF for high SNR, i.e., $\gamma\to \infty$, using the series expansion of the Meijer's G-function for small values, $\sqrt{\frac{\gamma_1^{\mathrm{thr}}}{c_1\gamma}}\to 0$,  \cite[Eq.~9.303 and Eq.~9.14.1]{TableIntegSerie8}. This approximation is given in (\ref{Eq:F1}) in Corollary~1.

\noindent
\textit{Case~2:} If $\gamma_1^{\mathrm{thr}}\gamma_2^{\mathrm{thr}}\geq1$ holds, $A$ is a non-zero constant which dominates $B$ for high SNR. In particular, $A$ can be written as $\mathrm{Pr}\big(\frac{1}{\gamma_2^{\mathrm{thr}}}<\frac{\gamma_{1}}{\gamma_{2}}<\gamma_1^{\mathrm{thr}}\big)$ which is given in terms of the CDF of the ratio of two G-G RVs, i.e., $\frac{\tilde{h}_1}{\tilde{h}_2}$, in (\ref{Eq:F2}) in Corollary~1 \cite[Eq. 7.821.3]{TableIntegSerie8}. This completes the proof.

\bibliographystyle{IEEEtran}
\bibliography{My_Citation_28-09-2017}

\end{document}